\documentclass[screen,acmsmall,nonacm]{acmart}
\AtBeginDocument{%
  }

\begin{document}

\title{Scientific Visualization as a Collaborative Data Infrastructure}

\author{Jasmine Tan Otto}
\correspondingauthor
\email{jtotto@ucsc.edu}
\affiliation{%
  \institution{University of San Francisco}
  \city{San Francisco}
  \state{California}
  \country{USA}
}

\author{Alexandra Diehl}
\affiliation{%
  \institution{University of Southern Denmark}
  \city{Odense}
  \country{Denmark}
}

\author{Max Kreminski}
\affiliation{%
 \institution{Cornell University}
 \city{New York}
 \state{New York}
 \country{USA}}

\author{Scott Davidoff}
\affiliation{%
  \institution{Space Science Institute}
  \city{Pittsburgh}
  \state{Pennsylvania}
  \country{USA}}

\renewcommand{\shortauthors}{Otto et al.}

\begin{abstract}
  Scientific visualization is an active site of infrastructuring with many layers of data and evidential claims. This paper reflects on collaborative Mars geoscience research conducted at the NASA Jet Propulsion Lab, which produced the PIXLISE spectroscopic analysis platform, through a retrospective analysis of a scientific discovery made by the team using PIXLISE. The history of infrastructuring in scientific visualization is exceedingly rich, which presents an opportunity for archival research. Simultaneously, the field of visualization aspires to become a science of communication, opening the door to future collaborations. Finally, with regard to visualization practitioners, we believe that opportunities for infrastructuring in both science and science communication are actively emerging.
\end{abstract}


\received{20 July 2026}

\maketitle

\section{Introduction}

While much of visualization research is framed around siloed, individual, cognitive activity~\cite{van_wijk_value_2005}, scientific research is in practice both deeply collaborative and remarkably heterogeneous~\cite{isenberg_collaborative_2011}.  Visualization practitioners collaborate with scientists to create tools that can help experts advance and critique scientific claims particular to an interdisciplinary project. Yet the handoff of `partial findings' between collaborators continues to be a key challenge in visual analytics research~\cite{zhao_supporting_2018}.

Drawing upon a partnership with NASA Jet Propulsion Lab geoscientists and their collaborators, who access data from the \emph{Perseverance rover} through the PIXLISE spectroscopy analysis platform~\cite{schurman:2019:pixelate}, we reflect on this shift from visualization systems toward infrastructure. This position paper parallels scientific visualization -- with its emphasis on building systems for data exploration and handoff -- with processes of infrastructuring in scientific collaboration more generally.

\begin{figure}
    \centering
    \includegraphics[width=1\linewidth]{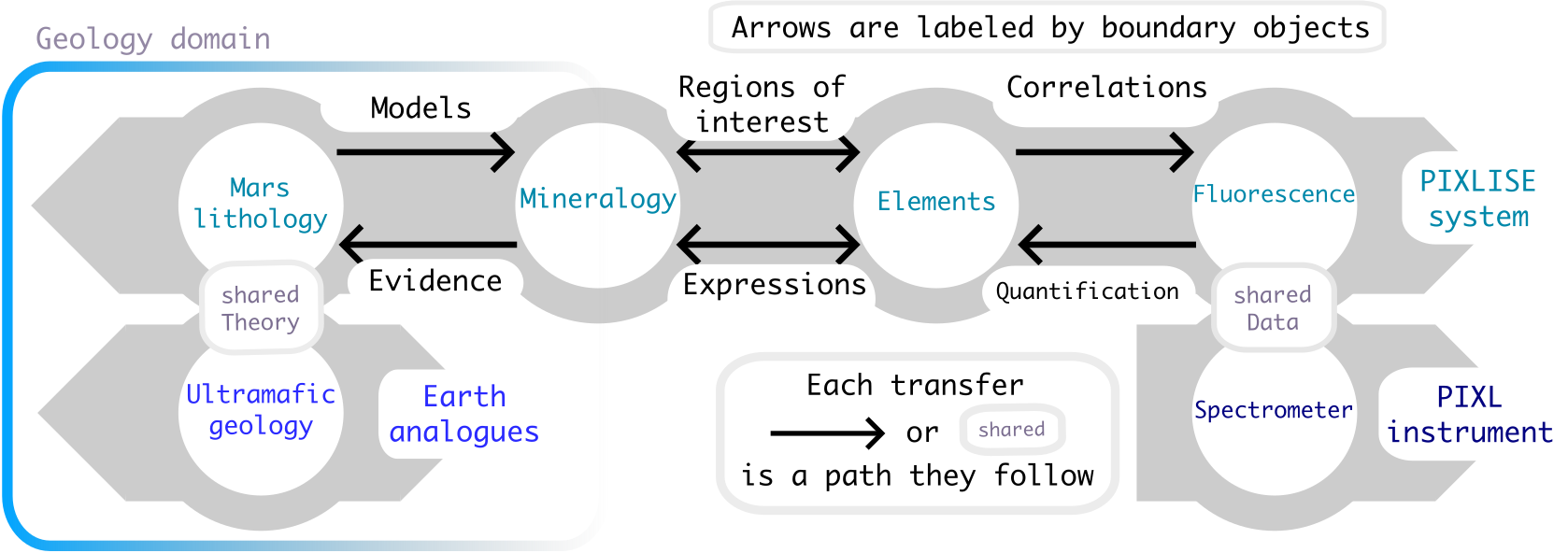}
    \caption{Knowledge transit diagram~\cite{otto_visualization_2024} depicting the flow of boundary objects (small black arrows) between `stations' (circles) of an infrastructure. From left to right, the stations proceed in decreasing order of abstraction, from geological theory to spectroscopic data gathered with the PIXL instrument (bottom-right). Two more `ground levels' are also visible: the implementation of PIXLISE itself (top-right), and the Earth analogues which inform scientists' lithological beliefs about Mars (bottom-left). This knowledge infrastructure enables rigorous scientific claims about Mars' 4.6 billion year history~\cite{liu_olivine_2022} to rest upon samples taken from a few square millimeters in Jezero crater on Mars.}
    \label{fig:transit}
\end{figure}

We develop two claims based on the PIXLISE case study:
\begin{itemize}
    \item P1: Successful collaborative sense-making interfaces incorporate many different \textit{boundary objects}~\cite{star_steps_1994}, which must be identified through stakeholders' existing practices.
    \item P2: Infrastructure builders must engage with multiple domains of knowledge~\cite{ribes_how_2019}, because \textit{handoff} regularly occurs across multiple levels (Fig.~\ref{fig:transit}).
\end{itemize}
We also discuss ways in which visualization research can motivate infrastructuring and vice versa.

\section{Background}
Visualization systems are used by experts to exchange findings and data across teams. Simple communication acts such as image embeds are supported by visualization components in messaging platforms. More specialized components include context images with a physical coordinate system; annotation layers thereon; and other ways of structuring complex datasets, in our case, fluorescence data gathered from Jezero crater on Mars by the \textit{Perserverence} rover.

In order to conduct Martian geoscience together, scientists, hardware experts, and visualization practitioners work across timezones and coordinate via their weekly cadence of meetings and share-outs. They work quickly when new Mars data arrive, because the knowledge gained from these experiments can be used to advocate for new experiments on subsequent days, and to send the rover to novel locations to conduct those experiments. Team members have diverse backgrounds and expertise, yet most stakeholders are not power users; they use PIXLISE to collect and view incoming data, but rarely use it to produce their own analyses.

\subsection{Sensemaking infrastructures support handoff}

Even when tightly designed for their domain-specific capacities, PIXLISE primarily serves experts' shared use. Often this is the first pass in their analyses. But in order to be satisfied why they are getting the readings that a collaborator has shared in PIXLISE, many scientists turn to other tools and theories that are outside its scope, such as Excel sheets of weight totals in a given region of interest. Despite being experts in their domains, and participating fully in discussions by \textbf{viewing shared representations} of the data produced from the visualization system, they often prefer to \textbf{explore the data} through their own established methods. They bring that knowledge back into science discussions, which are enriched and made more robust.

The visualization system is not a monolithic process. Rather, it connects boundary objects in active use by collaborators, such as geoscientists who use mixing diagrams (Fig.~\ref{fig:transit}). These objects already made up an informal infrastructure, which PIXLISE now operationalizes~\cite{beaudouin-lafon_generative_2021}.

\begin{figure}
    \centering
    \includegraphics[width=1\linewidth]{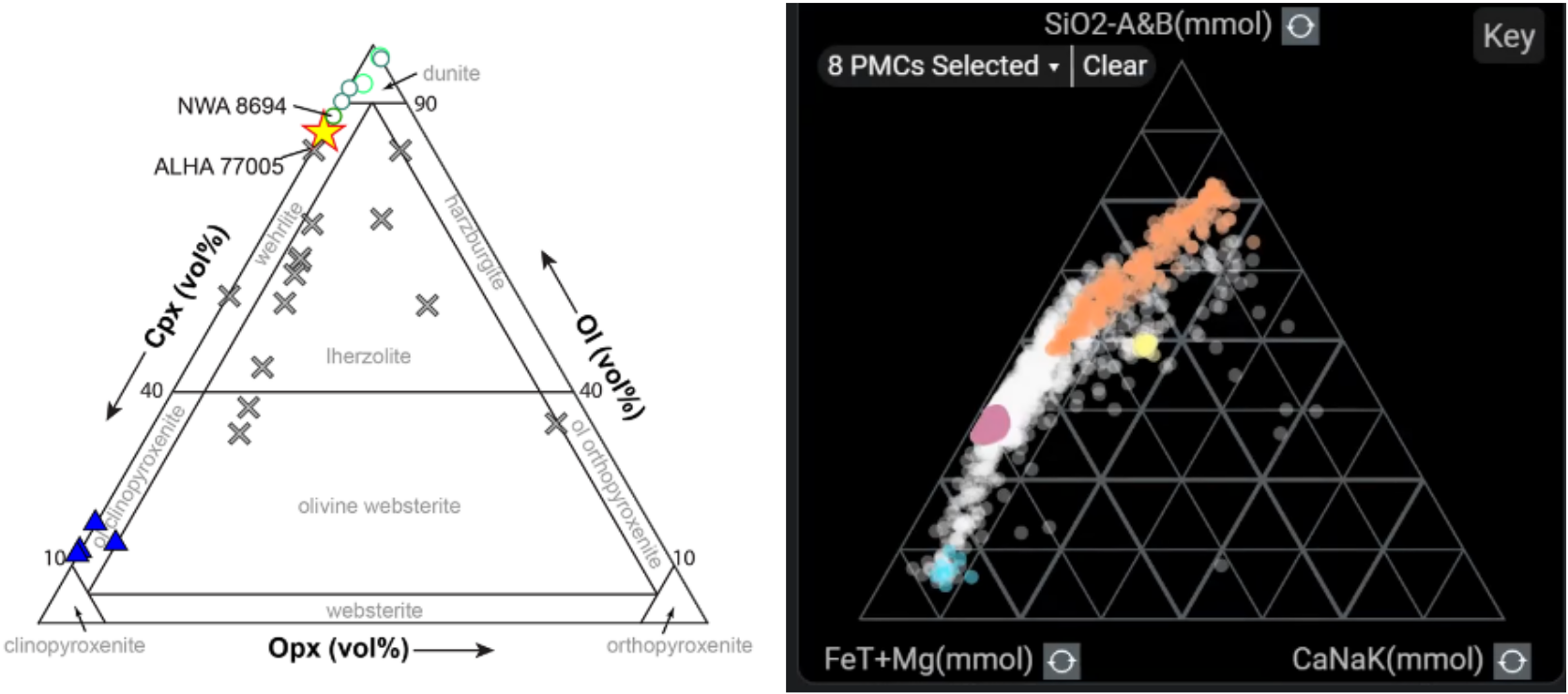}
    \caption{Representative examples of mixing diagrams in the geology literature and in PIXLISE, respectively. \textbf{Left:} Mixing diagram for ultramafic rock, reproduced from~\cite{liu_olivine_2022} (Fig.~3C), with the average of Dourbes pyroxene grains (sampled from Jezero crater) marked by a yellow star in the upper-left. `Ol' stands for olivine, `Cpx' for clinopyroxene, and `Opx' for orthopyroxene. \textbf{Right:} A mixing diagram in PIXLISE, containing sample points from Dourbes; color-coded by region of interest, with a user-driven selection visible in yellow. In PIXLISE, each axis corresponds to an \textit{expression}, e.g. here the upper vertex is quartz rather than olivine.}
    \label{fig:placeholder}
\end{figure}

\subsection{Visualization systems operationalize domain language}

In Fig.~\ref{fig:transit}, we identified four boundary objects specific to Mars geoscience practices which are operationalized in PIXLISE (\textbf{P1}). Two of these correspond roughly to specialized types of plot: the expressions, which form the axes of a mixing diagram; and the correlations, which form the $n^2$ edges of an $n$-element chord diagram. In the PIXLISE interface, mixing diagrams represent each sample point from the context map (a 2D representation of the sample patch in Jezero crater). Sample points in the mixing diagram (which cluster according to their chemical compostion) become selected when the corresponding physical location is `lassoed' on the context map and vice versa, employing the common linked brushing technique~\cite{isenberg_collaborative_2011}. 

Our two exceptions are `models' and `evidence', referring respectively to background knowledge and context drawn from scientific papers; and various materials produced using PIXLISE, especially screenshots and slide decks. These resources for scientific discussions are boundary objects that do not belong inside of PIXLISE, yet are boundary objects nonetheless; for the same reason that Lee and Schmidt have chosen to admit stakeholders' ad-hoc conversations as infrastructuring~\cite{lee_critical_2018}, even though they lack formality.

\section{Discussion}
Scientific visualization systems are a core example of infrastructuring with expert users to produce new insights from their data~\cite{van_wijk_value_2005}. This paper describes how the analysis needs of domain experts are identified, triaged, and communicated in the PIXLISE context (P1), and how PIXLISE itself supports handoff between experts with differing backgrounds (P2).

Scientific visualization systems regularly encounter the infrastructural tensions identified by Ribes and Finholt~\cite{ribes_long_2009}, despite high commitment from their users based at a variety of institutions. The broader visualization literature is also increasingly engaged with questions of rigor in situated research that reflect underlying infrastructuring needs, e.g. iteration, traceability, and reflection~\cite{rogers_reflections_2026}.

\subsection{Building for power users}

PIXLISE operationalizes the formal exchanges between geoscientists working together across different areas of expertise, along with spectroscopists and other collaborators. 

\begin{itemize}
    \item \textbf{Models} characterize which minerals should be present under a given Mars lithology.
    \item \textbf{Regions of interest} are spatial selections within the abrasion patch, e.g. grains of uniform composition, which are formed by certain lithological processes.
    \item \textbf{Correlations} between co-occurring elements in a region of interest. These are visualized in bulk by a \textbf{chord diagram}~(Fig.1d in \cite{ye_pixlise-c_2021}).
    \item \textbf{Quantification} of elements based on fluorescence data; for a given set of minerals, PIXLISE solves for a linear combination of elements which explains all observed spectral peaks.
    \item \textbf{Expressions} characterizing the proportions of different minerals based on available elements. Visualized in sets of three by \textbf{mixing diagrams}~(Fig.~\ref{fig:placeholder}).
    \item \textbf{Evidence} supporting a lithological claim (about the history of Mars) based on mineralogical evidence, e.g. the Dourbes site in Jezero crater. 
\end{itemize}

Each of these artifact types is produced from and illustrated by multiple maps or plots in the PIXLISE interface. While it remains possible to gather findings directly from reflectance data, the purpose of using PIXLISE is to take advantage of in-editor weight totals which can be cross-checked with elemental maps and the physical location of e.g. crystal grains in their rocky substrate.

The needs of power users emerge through their interaction with a tool, as they discover new use cases and push it past intended limits. In other words, the dataset used by a scientist does not constrain their use of novel methods and theoretical constructs; but if their tools do, then they will find a way to create the alternative they need. As such, scientific visualization includes a tradition of reincorporating these emergent use cases; in other words, \textit{desire paths}.

\subsection{Paving the desire paths}

PIXLISE was frequently extended to support previously ad-hoc analyses, turning the most structured part of the conversation into tools. Before regions of interest were integrated with expressions, teams would annotate a context image in Adobe Illustrator to produce maps of where the mineralogical features of interest were located. 

Exchanges of findings are conducted through internal chat platforms (Mattermost), e.g. by sharing a screenshot of some PIXLISE views. Team members would often share screenshots to get colleagues interested in their data. Someone else would then reproduce the analysis, usually not by performing literally the same steps in the same interface, but more often by using a different set of more familiar tools to inspect the underlying data.

So the analysis would be cross-checked through the constellation of visualization tools according to team members' differing expertise. Rather than boundary objects shaping collaboration dynamics, it is the conversations that shape the boundary objects.


\section{Conclusion}
Collaborative scientific visualization represents a rich set of case studies and relevant methodologies for data-intensive infrastructuring. Prototyping processes such as visualization design studies~\cite{crisan_passing_2021} pave the way for maintainable infrastructure; as visualization practitioners learn from power users' workflows, the boundary objects needed by other stakeholders become more crisply delineated and richly supported. On the Mars geoscience team, the source of new scientific knowledge is rarely a singular finding, but instead the consensus that emerges from contentious discussion and independent replication.

Visualization plays a special role in \textit{making common knowledge}, i.e. bringing technical arguments into conversations where they can help discern truths. Although space science is full of difficult problems, these are not wicked problems; it is clear that rocky formations visible on Mars arose through geological processes, and that competing theories of those processes are testable through physical sampling. Instead, the difficulty of visualization infrastructuring arises from the sheer number of domains in collaboration with each other. Geologists who study metamorphic rock rather than igneous rock hold different, conflicting contextual assumptions.

In our case study, comparisons between different elemental maps and mineralogical claims are facilitated by PIXLISE. More broadly speaking, institutions rely on boundary objects -- including reports, figures, and repositories of the aforementioned records, -- in order to know what they know. The diversity of boundary objects represents, translates, and transports common knowledge through infrastructure.  

The past four decades of scientific visualization research have, in the context of federal research institutions like NASA and INRIA, enabled dedicated scientists to produce historic results through interdisciplinary collaborations of extraordinary complexity. We believe this rich history will be of particular interest to infrastucture builders who want to recruit expert users, discover their unmet needs, and make new discoveries together.

\section*{Biography}

The authors are human-computer interaction researchers active in collaboration and visualization. They are interested in emerging infrastructures for public science and science communication.





\bibliographystyle{ACM-Reference-Format}
\bibliography{sample-base}

@String{Computer = "{IEEE} Computer" }

@inproceedings{schurman:2019:pixelate,
    author = {David Schurman and Pooja Nair and Scott Davidoff and Adrian Galvin and Abigail Allwood and Yang Liu and David Flannery and Robert P. Hodyss and Santiago V. Lombeyda and Maggie Hendrie and Hillary Mushkin and Christopher P. Heirwegh},
    title = {PIXELATE: Novel visualization and computational methods for the analysis of astrobiological spectroscopy data},
    booktitle={Proceedings of the 2019 Astrobiology Science Conference},
    series={AGU AbSciCon 2019},
    publisher={American Geophysical Union},
    year = {2019},
    url={https://agu.confex.com/agu/abscicon19/prelim.cgi/Paper/482995}
}

@misc{ye_pixlise-c_2021,
	title = {{PIXLISE}-{C}: {Exploring} {The} {Data} {Analysis} {Needs} of {NASA} {Scientists} for {Mineral} {Identification}},
	shorttitle = {{PIXLISE}-{C}},
	url = {http://arxiv.org/abs/2103.16060},
	doi = {10.48550/arXiv.2103.16060},
	language = {en},
	urldate = {2026-01-03},
	publisher = {arXiv},
	author = {Ye, Connie and Hermann, Lukas and Yildirim, Nur and Bhat, Shravya and Moritz, Dominik and Davidoff, Scott},
	month = mar,
	year = {2021},
	note = {arXiv:2103.16060 [cs]},
}

@article{liu_olivine_2022,
	title = {An olivine cumulate outcrop on the floor of {Jezero} crater, {Mars}},
	volume = {377},
	issn = {0036-8075, 1095-9203},
	url = {https://www.science.org/doi/10.1126/science.abo2756},
	doi = {10.1126/science.abo2756},
	language = {en},
	number = {6614},
	urldate = {2026-02-09},
	journal = {Science},
	author = {Liu et al., Y.},
	month = sep,
	year = {2022},
	pages = {1513--1519},
}

@article{ribes_long_2009,
	title = {The {Long} {Now} of {Technology} {Infrastructure}: {Articulating} {Tensions} in {Development}},
	volume = {10},
	issn = {1536-9323},
	shorttitle = {The {Long} {Now} of {Technology} {Infrastructure}},
	url = {https://aisel.aisnet.org/jais/vol10/iss5/5},
	doi = {10.17705/1jais.00199},
	number = {5},
	journal = {Journal of the Association for Information Systems},
	author = {Ribes, David and Finholt, Thomas},
	month = may,
	year = {2009},
}

@article{crisan_passing_2021,
	title = {Passing the {Data} {Baton} : {A} {Retrospective} {Analysis} on {Data} {Science} {Work} and {Workers}},
	volume = {27},
	copyright = {https://ieeexplore.ieee.org/Xplorehelp/downloads/license-information/IEEE.html},
	issn = {1077-2626, 1941-0506, 2160-9306},
	shorttitle = {Passing the {Data} {Baton}},
	url = {https://ieeexplore.ieee.org/document/9222030/},
	doi = {10.1109/TVCG.2020.3030340},
	language = {en},
	number = {2},
	urldate = {2026-07-21},
	journal = {IEEE Transactions on Visualization and Computer Graphics},
	author = {Crisan, Anamaria and Fiore-Gartland, Brittany and Tory, Melanie},
	month = feb,
	year = {2021},
	pages = {1860--1870},
}

@inproceedings{otto_visualization_2024,
	address = {St Pete Beach, FL, USA},
	title = {Visualization {Artifacts} are {Boundary} {Objects}},
	copyright = {https://doi.org/10.15223/policy-029},
	isbn = {979-8-3315-2846-1},
	url = {https://ieeexplore.ieee.org/document/10756169/},
	doi = {10.1109/BELIV64461.2024.00014},
	language = {en},
	urldate = {2026-07-21},
	booktitle = {2024 {IEEE} {Evaluation} and {Beyond} - {Methodological} {Approaches} for {Visualization} ({BELIV})},
	publisher = {IEEE},
	author = {Otto, Jasmine T. and Davidoff, Scott},
	month = oct,
	year = {2024},
	pages = {81--88},
}

@article{ribes_how_2019,
	title = {How {I} {Learned} {What} a {Domain} {Was}},
	volume = {3},
	issn = {2573-0142},
	url = {https://dl.acm.org/doi/10.1145/3359140},
	doi = {10.1145/3359140},
	language = {en},
	number = {CSCW},
	urldate = {2026-07-21},
	journal = {Proceedings of the ACM on Human-Computer Interaction},
	author = {Ribes, David},
	month = nov,
	year = {2019},
	pages = {1--12},
}

@inproceedings{star_steps_1994,
	address = {Chapel Hill, North Carolina, United States},
	title = {Steps towards an ecology of infrastructure: complex problems in design and access for large-scale collaborative systems},
	copyright = {https://www.acm.org/publications/policies/copyright\_policy\#Background},
	isbn = {978-0-89791-689-9},
	shorttitle = {Steps towards an ecology of infrastructure},
	url = {http://portal.acm.org/citation.cfm?doid=192844.193021},
	doi = {10.1145/192844.193021},
	language = {en},
	urldate = {2026-07-21},
	booktitle = {Proceedings of the 1994 {ACM} conference on {Computer} supported cooperative work  - {CSCW} '94},
	publisher = {ACM Press},
	author = {Star, Susan Leigh and Ruhleder, Karen},
	year = {1994},
	pages = {253--264},
}

@misc{rogers_reflections_2026,
	title = {Reflections on {Traceability} for {Visualization} {Research}},
	url = {http://arxiv.org/abs/2604.14417},
	doi = {10.1111/cgf.70445},
	language = {en},
	urldate = {2026-04-20},
	author = {Rogers, Jen and Akbaba, Derya and Scott-Brown, James and Lex, Alexander and Meyer, Miriah},
	month = apr,
	year = {2026},
	note = {arXiv:2604.14417 [cs]},
}

@article{isenberg_collaborative_2011,
	title = {Collaborative visualization: {Definition}, challenges, and research agenda},
	volume = {10},
	issn = {1473-8716},
	shorttitle = {Collaborative visualization},
	url = {https://doi.org/10.1177/1473871611412817},
	doi = {10.1177/1473871611412817},
	language = {EN},
	number = {4},
	urldate = {2026-06-03},
	journal = {Information Visualization},
	publisher = {SAGE Publications},
	author = {Isenberg, Petra and Elmqvist, Niklas and Scholtz, Jean and Cernea, Daniel and Ma, Kwan-Liu and Hagen, Hans},
	month = oct,
	year = {2011},
	pages = {310--326},
}

@incollection{lee_critical_2018,
	address = {Oxford},
	title = {Critical remarks on the concept of ‘infrastructure’ in {CSCW} and {IS}},
	language = {en},
	booktitle = {Socio-{Informatics}: {A} {Practice}-based {Perspective} on the {Design} and {Use} of {IT} {Artifacts}},
	publisher = {Oxford University Press},
	author = {Lee, Charlotte P and Schmidt, Kjeld},
	year = {2018},
	pages = {177--217 (chapter 5)},
}

@article{zhao_supporting_2018,
	title = {Supporting {Handoff} in {Asynchronous} {Collaborative} {Sensemaking} {Using} {Knowledge}-{Transfer} {Graphs}},
	volume = {24},
	copyright = {https://ieeexplore.ieee.org/Xplorehelp/downloads/license-information/IEEE.html},
	issn = {1077-2626},
	url = {http://ieeexplore.ieee.org/document/8017596/},
	doi = {10.1109/TVCG.2017.2745279},
	language = {en},
	number = {1},
	urldate = {2026-07-21},
	journal = {IEEE Transactions on Visualization and Computer Graphics},
	author = {Zhao, Jian and Glueck, Michael and Isenberg, Petra and Chevalier, Fanny and Khan, Azam},
	month = jan,
	year = {2018},
	pages = {340--350},
}

@inproceedings{van_wijk_value_2005,
	title = {The value of visualization},
	url = {https://ieeexplore.ieee.org/abstract/document/1532781},
	doi = {10.1109/VISUAL.2005.1532781},
	urldate = {2026-07-21},
	booktitle = {{VIS} 05. {IEEE} {Visualization}, 2005.},
	author = {van Wijk, J.J.},
	month = oct,
	year = {2005},
	pages = {79--86},
}

@article{beaudouin-lafon_generative_2021,
	title = {Generative {Theories} of {Interaction}},
	volume = {28},
	issn = {1073-0516, 1557-7325},
	url = {https://dl.acm.org/doi/10.1145/3468505},
	doi = {10.1145/3468505},
	language = {en},
	number = {6},
	urldate = {2025-10-29},
	journal = {ACM Transactions on Computer-Human Interaction},
	author = {Beaudouin-Lafon, Michel and Bødker, Susanne and Mackay, Wendy E.},
	month = dec,
	year = {2021},
	pages = {1--54},
}

\end{document}